# HIGH VOLTAGE GENERATION FOR PHYSICS LAB

RAJU BADDI
National Center for Radio Astrophysics, TIFR, Ganeshkhind P.O.Bag 3,
Pune University Campus, PUNE 411007, MAHARASHTRA, INDIA;
baddi@ncra.tifr.res.in

## ABSTRACT

A power efficient way to generate low power high voltage is given. The article describes various aspects of functioning and derives quantitative relations between different parameters and high voltage generated. Use of voltage multiplier(Cockcroft-Walton multiplier) network can provide further boost in the high voltage(~1000V).

## 1. INTRODUCTION

Higher voltage from a lower one can be obtained using a step-up transformer(Kasatkin & Nemtsov 1986). Here such a circuit is analysed and quantitative relations between various parameters are derived for the sake of easy tailoring of the circuit as per ones requirement. It consists of a pulsative current drive circuit for the primary of the transformer using a transistor switch and a rectangular wave oscillator. The output of the secondary which provides the high voltage pulses can be connected to a bridge rectifier and a capacitor filter as shown in Figure 1 or further enhancement in voltage can be achieved by using a voltage multiplier network(Cockcroft-Walton multiplier).

## 2. THE HIGH VOLTAGE GENERATOR

This article describes a simple, efficient way to generate high voltage using a step-up transformer preferably with a ferrite core. Analysis of the circuit is made so that the electronics and transformer windings could be tailored to specific needs. The step up transformer consists of two coils, one with a lower number of turns(N1, $L_1$, primary) and the other with a higher number of turns(N2, $L_2$, secondary). The coil with N1 turns carries a higher current while the coil with N2 turns gives a higher voltage but carries a much lower current. We will call the N1 as primary winding and N2 as the secondary winding. The plan is to turn on a relatively heavy varying current in N1 at lower voltage and obtain a much higher voltage in N2 at the cost of reduced current. The current in the primary is switched on and off intermitently at a specific frequency with the help of a power transistor Q1. The transistor is operated by a power efficient CMOS oscillator formed by gates G1,G2 and G3||G4. The schematic circuit diagram is shown in Figure 1. Q1 is turned on when the output of the gates G3(G4) is low. G3 & G4 have been wired in parallel to deliver more current to the base of Q1. It is also possible to put transistors in parallel to handle more current. Driving each base with an independent gate output through a base resistor and having common emitter/ collector terminals. The primary($L_1$) is assumed to be made of sufficiently thick copper wire so that it can handle heavy currents from 10's of mAs to 100's of mAs with negligible resistive voltage drop across it. This also ensures a linear increase in current in the primary which results in a constant voltage at the secondary. Added to this there will be no loss of power in the primary due to heating. Q1 as should be noted plays the role of an electronic switch. While choosing Q1 one has to pick a pnp power transistor with good switching characteristics and required current handling capabilities. When Q1 is switched on (called $t_{on}$ period)it results in a constant voltage across the primary whose inductance we will denote as $L_1$. The constant voltage results in the generation of linearily rising current in the primary. Hence the magnetic field in the core rises linearily. This induces a constant high voltage in secondary($L_2$) due to its larger number of turns which couple with the magnetic field. The strength of the current increases steadily in the primary builing up energy in the magnetic field of the core. This magnetic field in the core couples to both the coils and hence can transfer energy from the battery into the secondary coil or it acts as a temporary reservoir of energy to which both the coils add/withdraw energy. Basically the primary pumps energy into the magnetic field of core while the secondary draws energy from the magnetic field and dissipiates it in the load resistance $R_L$ (not shown in Figure 1 but connected to HV+ and HV-). The primary converts the electrical energy of the battery into magnetic energy while the secondary converts this magnetic energy back into electrical energy but just at a different voltage with the same power. The appendix gives the simplified details of the dependance of the secondary voltage as a function of different circuit parameters. These equations have been verified under computer simulations satisfactorily over a range of frequencies and other parameters.

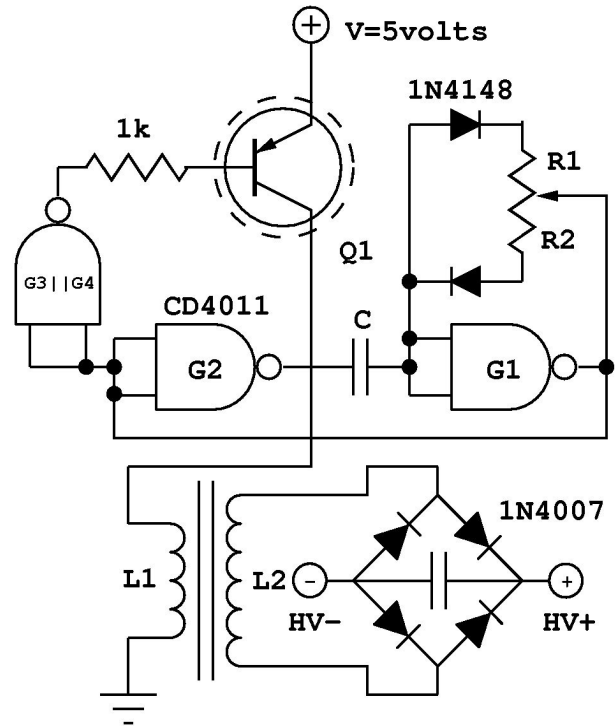

*Fig 1: Typcial example schematic circuit diagram of High Voltage Generator. In all discussion Q1 is assumed to play the role of an ideal electronic switch. Oscillator starting-problem issues if any can be remedied refering Design Ideas, Apr 21 2011, **EDN**.*

Typical examples for Q1 are BC177,2N3467,ZTX749,ZTX550,ZTX788B.

$L_1$ = 100μ-5mH.

$L_2$ = 10mH – 5H.

$t_{on}/t_{off}$ = 5μs to 100μs

The gate time periods of logic high are as under,

$$t^{G1}_{high} \approx R_1 C \ln\left(\frac{V+V_T-0.6}{V_T-0.6}\right) \;;\; t^{G2}_{high} \approx R_2 C \ln\left(\frac{2V-V_T-0.6}{V-V_T-0.6}\right) \quad (1)$$

$V_T$ is the threshold voltage of the gate. From these $t_{on}$ and $t_{off}$ can be set to desired values by picking suitable values of $R_1,R_2$ and C. (1) assumes gate G1's inputs to be completely non-sourcing/sinking current for any input voltage which is not true. However incorporating a high resistance(~100KΩ) before its common inputs can make (1) much more reliable.

Important results of the appendix are,

$$V_2^{+mx} = \sqrt{\frac{L_2}{L_1}} V \;;\; V_2^{-mx} = \frac{V t_{on} R_L}{(1-\xi)\sqrt{L_1 L_2}} \;;\; \xi = e^{\frac{-R_L}{L_2}t_{off}} \quad (2)$$

Where $V_2^{+mx}$ and $V_2^{-mx}$ are the maximum positive and negative peaks at the terminals of $L_2$ (without the rectifier bridge/capacitor). It should be noted that while $V_2^{+mx}$ is constant over $t_{on}$, $V_2^{-mx}$ is not over $t_{off}$(Figure 2). The implementation of the circuit is simple. Once the transformer windings are fabricated one has to simply measure inductances of primary and secondary independently. These values can then be used in (2) to obtain the voltages induced in the secondary. It should be noted that voltages in (2) are across $R_L$ directly connected to the secondary. However during generation of large voltages the voltage drop across the bridge diodes can be neglected. The capacitor included in the bridge is to smooth out the varying voltages. Further it is possible to have nearly symmetric +ve/-ve impulses of equal magnitude by choosing appropriate parameters. It should be noted that (1) is based on Author's derivations. The appendix details the derivation of (2).

# APPENDIX

First we write the equations for the time period $t_{on}$ for which the transistor is turned on. This happens on the falling edge of gates G3||G4. Since the primary and secondary coils are on the same core we have their mutual inductance $M_{12}=M_{21}=M=k\sqrt{L_1 L_2}=\sqrt{L_1 L_2}$ (coupling coefficient, *k* taken to be 1). So the equation for the primary coil is,

$$L_1 \frac{di_1}{dt} + M\frac{di_2}{dt} = V \qquad (A1)$$

Similarly we write for the secondary as,

$$L_2 \frac{di_2}{dt} + M\frac{di_1}{dt} + i_2 R_L = 0 \qquad (A2)$$

Where $R_L$ is the load resistance connected to HV+/HV- neglecting the diode voltage drop, or in other words $R_L$ is directly connected across the secondary without the bridge rectifier in Figure 1. Using (A1) to eliminate $di_1/dt$ in (A2) results in,

$$L_2 \frac{di_2}{dt} + M\left(\frac{V}{L_1} - \frac{M}{L_1}\frac{di_2}{dt}\right) + i_2 R_L = 0 \qquad (A3)$$

which gives $\qquad i_2 = -\frac{MV}{L_1 R_L} \quad ; \quad \frac{di_1}{dt} = -\frac{V}{L_1} \qquad (A4)$

$i_2$ is the steady current that flows through $R_L$ once the supply voltage V is established across the primary. However as the transistor Q1 is turned on/off intermittently we now consider a situation when Q1 is turned on after an off state. It should be noted that during off state we assume the current through the primary coil to be zero. So the magnetic field which had been feeding energy into the secondary coil by its decay has reduced in magnitude during the time $t_{off}$. The current in the secondary coil determines the magnetic field strength completely as there is no other current(here primary) to sustain the magnetic field. When Q1 is turned on, due to absence of resistance in primary the sustainence of magnetic field is rapidly taken over by the primary current. Very rapidly an equivalent current required to sustain the existing magnetic field in the core is setup in the primary and the current in the secondary also ceases at the same pace, essentially energy is conserved. Additionally due to increase in the current in the primary a steady current or voltage is established in secondary during $t_{on}$(equation A4, Figure 2)). Since the currents in primary and secondary are in opposite directions the magnetic fields produced by the two coils in the core are in opposite directions. These respective currents are also established very rapidly without any hinderance. To start with we understand first what an equivalent current means when a current from one coil is immediately transfered to another coil for the sake of magnetic field sustainance. If a current $i_1$ flowing through $L_1$ can produce a magnetic field B in the core then a current $i_2$ can also produce the same magnetic field B in the core. These currents using the energy conservation $L_1 i_1^2 = L_2 i_2^2$ can be written as,

$$i_1 = \sqrt{\frac{L_2}{L_1}} i_2 \quad ; \quad i_2 = \sqrt{\frac{L_1}{L_2}} i_1 \qquad (A5)$$

We write for the current in primary during $t_{on}$ as,

$$i_1(t) = \left[ i_r + \frac{L_2 V}{L_1 R_L} \right] + \frac{V}{L_1} t \qquad (A6)$$

The current in the brackets in (A6) is immediately established in the primary when $t_{on}$ starts. $i_r$ is the residual current in the secondary that is transfered to the primary where as the second term in the bracket is due to the steady current or voltage in the secondary, equation (A4). $i_2$ in (A4) and this term are opposite to each other and contribute together nothing to the core magnetic flux and hence are easily established without any hinderance. We now write $i_r$ as the magnitude of the decaying current in secondary at the end of $t_{off}$ as,

$$i_r = e^{\frac{-R_L}{L_2} t_{off}} \sqrt{\frac{L_2}{L_1}} \; i_2^{mx} = i_2^{mx} \xi \sqrt{\frac{L_2}{L_1}} \; . \qquad (A7)$$

Note the substitution $\xi$. At the end of $t_{on}$ equation (A6) gives the strength of current in the primary,

$$i_1(t_{on}) = i_1^{mx} = \left[ i_r + \frac{L_2 V}{L_1 R_L} \right] + \frac{V}{L_1} t_{on} \qquad (A8)$$

At the end of $t_{on}$ Q1 is switched off and we assume that the current in the primary immediately ceases to exist. But this does not mean that the magnetic flux vanishes in the core. At this moment the secondary takes charge of the magnetic flux and sustains it by its current. However due to its large inductance and a large resistance $R_L$ in its circuit the voltage required in the secondary to sustain this magnetic field would be large. The decaying magnetic field produces this. We have for the immediate current in the secondary at the start of $t_{off}$ as,

$$i_2^{mx} = \sqrt{\frac{L_1}{L_2}} \; i_1^{mx} - \sqrt{\frac{L_2}{L_1}} \; \frac{V}{R_L} \qquad (A9)$$

where we have used the transfer relations (A5). The second term on the right hand side is the steady current due to the ramping current $di_1/dt$ during $t_{on}$, equation (A4). Substituting for $i_1^{mx}$ from (A8) we obtain the maximum voltage i.e the peak -ve impulse in the secondary during the decay of magnetic field in the core. This after simplification is as given under,

$$i_2^{-mx} = \frac{V t_{on}}{(1-\xi)\sqrt{L_1 L_2}} \quad ; \quad V_2^{-mx} = \frac{V t_{on} R_L}{(1-\xi)\sqrt{L_1 L_2}} \tag{A10}$$

The +ve voltage is the steady or constant voltage due to the ramping current $di_1/dt$ during $t_{on}$ of primary as given in equation (A4) and is given as under combined with $V_2^{-mx}$,

$$V_2^{+mx} = \sqrt{\frac{L_2}{L_1}} V \quad ; \quad V_2^{-mx} = \frac{V t_{on} R_L}{(1-\xi)\sqrt{L_1 L_2}} \tag{A11}$$

It should be noted that where as $V_2^{+mx}$ is constant during $t_{on}$ $V_2^{-mx}$ changes/reduces in magnitude exponentially during $t_{off}$ according to the R-L circuit as given under(see Figure 2),

$$V_2^{-ve}(t) = \frac{V t_{on} R_L}{(1-\xi)\sqrt{L_1 L_2}} e^{\frac{-R_L}{L_2} t} \tag{A12}$$

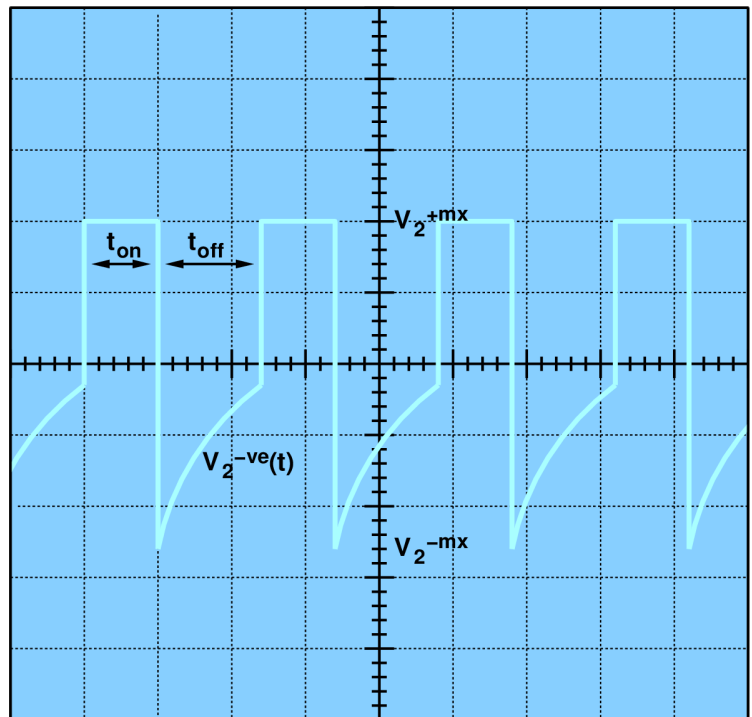

*Fig 2: Typical pulse waveform at the terminals of $L_2$.*